# Low-pressure CVD grown Si-doped β-Ga$_2$O$_3$ films with promising electron mobilities and high growth rates

Saleh Ahmed Khan[1], Ahmed Ibreljic[1], Stephen Margiotta[1], A F M Anhar Uddin Bhuiyan[1, a)]

[1]Department of Electrical and Computer Engineering, University of Massachusetts Lowell, Lowell, MA 01854, USA

a) Corresponding author Email: anhar_bhuiyan@uml.edu

**Abstract**

In this work, we systematically investigated the growth of Si-doped β-Ga$_2$O$_3$ films using Low pressure chemical vapor deposition (LPCVD) system, achieving high room temperature (RT) hall mobilities of 162 cm²/V·s and 149 cm²/V·s at carrier concentrations of $1.51 \times 10^{17}$ cm$^{-3}$ and $1.15 \times 10^{17}$ cm$^{-3}$, respectively for homoepitaxial (010) β-Ga$_2$O$_3$ film grown on β-Ga$_2$O$_3$ substrates and heteroepitaxial ($\bar{2}$01) β-Ga$_2$O$_3$ film grown on off-axis c-sapphire substrate with 6° miscut-representing the highest mobilities reported for LPCVD-grown β-Ga$_2$O$_3$ materials. Carrier concentrations were precisely tuned by varying SiCl$_4$ flow rates at growth temperature of 1000 °C, resulting in concentrations ranging from $1.15 \times 10^{17}$ to $1.19 \times 10^{19}$ cm$^{-3}$ as confirmed by both Hall and capacitance-voltage (C-V) measurements. The films exhibited high crystalline quality, confirmed by high resolution X-ray diffraction (XRD) and Raman spectroscopy, indicating phase purity and structural integrity. Surface morphologies characterized by field-emission scanning electron microscope (FESEM) and atomic force microscopy (AFM) showed a strong correlation between carrier concentrations and surface smoothness, with lower concentration resulting in reduced RMS roughness. Secondary Ion Mass Spectrometry (SIMS) analysis revealed uniform Si incorporation, with low carbon, hydrogen, and chlorine impurities below detection limits, indicating high purity of the films. A high low-temperature peak mobility exceeding >843 cm²/V·s was achieved for ($\bar{2}$01) β-Ga$_2$O$_3$ heteroepitaxial films at 80 K, highlighting the high purity and low compensation of these films. These findings emphasize the potential of LPCVD growth system for producing high-purity β-$Ga_2O_3$ films with thickness ranging between ~2.3-11.7 μm and faster growth rates (~4.7-17 μm/hr), promising transport properties, controllable doping, and scalability for developing high power vertical devices.

As the demand for advanced power-switching devices intensifies with the global shift toward electrification, semiconductor materials capable of operating under high voltage are becoming increasingly vital. While wide-bandgap materials such as SiC and GaN have surpassed Si for high-power applications, ultra-wide bandgap materials, particularly β-$Ga_2O_3$ are now gaining prominence as leading candidates for the next generation of high-performance power electronics due to its ultra-wide bandgap energy (~4.8 eV), placing it in the deep ultraviolet spectrum. The controllable n-type doping and high breakdown field strength of β-$Ga_2O_3$, projected to reach up to 8 MV/cm, surpasses that of SiC and GaN, making it a prime candidate for high voltage switching applications [1]. Moreover, β-$Ga_2O_3$ is unique in its ability to be grown by melt-based growth methods [2], offering scalability and cost-effectiveness through larger substrate sizes, enabling significant advancements in developing high power devices. In recent years, significant progress in β-$Ga_2O_3$-based power diodes and transistors have achieved multi-KV breakdown voltages, making them well-suited for applications such as electric vehicles, power grid, renewable energy, and defense. Vertical device architectures, in particular, are highly desirable for high-power applications as they allow for superior current drive and efficient field termination [3-18]. Achieving thick drift layers with low background carrier concentration is crucial for the scalability and performance of these devices. While several epitaxial growth techniques have been employed to grow β-Ga$_2$O$_3$ films, such as hydride vapor phase epitaxy (HVPE) [19-21], metalorganic chemical vapor deposition (MOCVD) [22-33], LPCVD [34-40], and molecular beam epitaxy (MBE) [41-48], each has its own limitations. HVPE, for instance, is known for its rapid growth rates (~10-250 µm/h), making it a dominant choice for vertical device demonstrations. However, HVPE's fast growth rates often result in surface roughness that requires chemical-mechanical polishing before device fabrication. Recent advances in MBE growth technique [44-48], particularly suboxide and hybrid plasma-assisted MBE, has demonstrated controllable Si doping of β-Ga$_2$O$_3$ with promising mobilities [44,46] and higher growth rates (~1 µm/hr) [45]. Among these techniques, MOCVD is considered as a highly promising method for growing β-Ga$_2$O$_3$ films due to its ability to produce superior crystalline quality and controlled doping. MOCVD growth using triethylgallium (TEGa) and trimethylgallium (TMGa) has demonstrated record-high electron mobilities with low doping concentrations [22-24, 30]. However, MOCVD faces challenges in achieving higher growth rates, especially with TEGa, while TMGa allows for higher growth rates but introduces issues such as surface defects and high carbon incorporation, degrading material purity, surface quality, and transport properties [24]. In contrast, LPCVD offers several advantages for the growth of β-Ga$_2$O$_3$ films. LPCVD provides precise control over doping concentrations, minimizes impurity incorporation, and enables smoother, more uniform surfaces with much higher growth rates (>10 µm/h) [34-40]. This is particularly advantageous for high-power applications, where thicker drift layer, surface quality, background impurities and doping precision are critical. Unlike MOCVD, which relies on metal-organic precursors susceptible to introducing unwanted impurities, LPCVD uses metallic Ga as a precursor, potentially enabling higher material purity and improved electron mobility. The controlled low-pressure environment also facilitates the deposition of thick drift layers with minimal defects [39]. LPCVD eliminates the use of organic ligands, preventing the risk of carbon contamination and enabling a faster decomposition process, resulting in higher growth rates, improved lattice stability with less defects, while the low-pressure environment ensures uniform film growth by improving precursor adsorption and minimizing diffusion barriers. Numerical simulation studies, such as those using computational fluid dynamics [34], have been explored to analyze gas-phase reactions and mass transport limitations in LPCVD growth systems. Additionally, advanced simulation techniques such as density functional theory (DFT) and mesoscopic phase-field modeling [49, 50] provide a theoretical framework that could complement experimental efforts by offering insights into factors such as gas-phase dynamics, temperature gradients, and defect formation during LPCVD growth. Previous experimental studies on LPCVD-grown β-$Ga_2O_3$ have shown great potential in achieving high quality films [34-40]. In this current study, we seek to further explore LPCVD systems to push the current boundaries of electrical transport and structural quality of both homo- and hetero-epitaxial β-$Ga_2O_3$ films with a goal of achieving superior electron mobility while maintaining decent surface morphologies with faster growth rates. By fine-tuning LPCVD process parameters- such as temperature, gas flow rates, and source-substrate distances, high-quality films with promising electron mobilities and controllable doping was achieved, highlighting the potential of LPCVD system for the scalable production of thick β-$Ga_2O_3$ films for developing high power vertical devices.

The β-$Ga_2O_3$ growths have been performed in a custom built LPCVD system on off-axis c-sapphire substrates with 6˚ offcut and (010) β-$Ga_2O_3$ Fe-doped substrates. All the samples were cleaned using acetone, IPA and DI water followed by $N_2$ drying. Argon (99.9999% purity) has been utilized as the neutral carrier and purge gas; oxygen (99.999% purity) and gallium pellets (99.99999% purity) were used as O- and Ga- source precursors, respectively. $SiCl_4$ (flow rates: 0.03 to 0.5 sccm) has been used as the n-type dopant source. The growths were performed with Ar and $O_2$ flows of 200 and 20 sccm, respectively. The growth temperature was 1000°C and the growth pressure was ~1.5 Torr. Ga source-to-substrate distances were tuned between 1.5 and 3.5 cm. The surface morphologies were characterized using a JEOL JSM 7401F FESEM and a Park XE-100 AFM systems. The thickness of homoepitaxial films were estimated from SIMS and FESEM cross-sectional images of the coloaded films grown on c-plane sapphire. The crystalline structure and quality were evaluated using XRD, performed with a Rigaku SmartLab tool equipped with a Cu Kα radiation source (λ=1.5418 Å). The Raman spectroscopy has been carried out utilizing a Horiba LabRam Evolution Multiline Raman Spectrometer (λ = 532 nm). Transport properties were measured using van der Pauw-Hall configuration using the Ecopia HMS-5300 Hall effect measurement system from room to low temperatures (80 K). In order to measure transport properties by Hall measurements, 30/100 nm Ti/Au ohmic contacts were deposited on the four corners of each homoepitaxial β-$Ga_2O_3$ film. Vertical Schottky barrier diode structures using LPCVD homoepitaxial films grown on Sn-doped (010) β-$Ga_2O_3$ substrates were fabricated for C-V measurements. Device fabrication began with the deposition of a Ti/Au (30/100 nm) Ohmic metal stack on the backside of the substrates via E-beam evaporation, followed by rapid thermal in a nitrogen atmosphere at 470°C for 1 minute. Subsequently, Schottky C-V pads were lithographically defined, and a Schottky metal stack of Ni/Au (20/100 nm) was deposited on top of β-$Ga_2O_3$ layer. The C-V measurements were conducted using a Keithly 4200A parameter analyzer.

The FESEM images of the surface morphologies of LPCVD-grown β-$Ga_2O_3$ films on (010) β-$Ga_2O_3$ native substrates and off-axis c-sapphire substrates with a 6° offcut are shown in Figure 1. Figures 1(a-c) show the homoepitaxial films grown with different carrier concentrations, ranging from $1.51 \times 10^{17}$ to $1.19 \times 10^{19}$. All samples exhibit a diagonal terrace-like morphology with multiple steps, consistent with previous reports on LPCVD-grown β-Ga$_2$O$_3$ films [35-37]. A strong relationship between surface morphology and carrier concentrations can be observed: as the dopant incorporation decrease, the surface becomes smoother, with smaller features, which is also accompanied by an increase in electron mobility. Figures 1(d-f) show heteroepitaxial ($\bar{2}$01) β-Ga$_2$O$_3$ film surfaces grown on 6° offcut c-plane sapphire, where step-flow growth along the offcut is evident. A similar trend is observed- higher doping concentrations lead to rougher surface morphology. To further investigate the surface roughness, AFM imaging was performed. Figures 2(a) and (b) show AFM scans (5 × 5 μm²) of homoepitaxial β-Ga$_2$O$_3$ films with doping concentrations of $1.51 \times 10^{17}$ cm$^{-3}$ and $1.39 \times 10^{18}$ cm$^{-3}$, yielding RMS roughness values of 2.65 nm (film thickness: 3.40 μm) and 3.26 nm (film thickness: 3.37 μm), respectively, comparable to previously reported LPCVD-grown films [35-37]. Figure 2(c) and (d) presents the AFM scans of heteroepitaxial β-Ga$_2$O$_3$ film surfaces with doping concentrations of $2.27 \times 10^{17}$ cm$^{-3}$ and $1.46 \times 10^{18}$ cm$^{-3}$, revealing clear and distinguishable steps along the offcut and an RMS roughness of ranging between 4.29 and 5.14 nm, measured over a 2 × 2 μm² area.

To investigate the crystalline structure and quality of the films, XRD ω-2θ and ω-rocking curve scans were performed, as shown in Figures 3(a) and (b) for β-Ga$_2$O$_3$ homoepitaxial and heteroepitaxial films, respectively, with varying doping concentrations. The wide-range ω-2θ scan in the inset of Figure 3(a) reveals a distinct, high-intensity (020) β-Ga$_2$O$_3$ peak for the homoepitaxial film, while both ($\bar{4}$02) and ($\bar{8}$04) peaks from ($\bar{2}$01) β-Ga$_2$O$_3$ films are observed in the inset of Figure 3(b) for the heteroepitaxial growth on c-sapphire with a 6° miscut. No additional peaks associated with other planes and phases of Ga$_2$O$_3$ (α, γ, δ, ε) were detected, indicating the growth of single-crystal β-Ga$_2$O$_3$ films. Lower rocking curve FWHM values of 67, 63, and 124 arc-sec were observed for the homoepitaxial films with doping concentrations of $1.51 \times 10^{17}$, 2.25

× 10$^{18}$, and 1.19 × 10$^{19}$ cm$^{−3}$ respectively, highlighting the high crystalline quality of the films. Figure 3(b) shows the rocking curves for the heteroepitaxial ($\bar{2}$01) β-Ga$_2$O$_3$ films, with smaller FWHMs of 293.04, 292.68, and 297.72 arc-sec for doping concentrations ranging from 1.15 × 10$^{17}$ to 8.29 × 10$^{18}$ cm$^{−3}$, demonstrating superior crystalline quality compared to previously reported LPCVD, MBE, MOCVD-grown heteroepitaxial thin films [36, 37, 51-53].

In addition, both homo- and hetero-epitaxial β-Ga$_2$O$_3$ thin films were characterized using Raman spectroscopy to investigate their phonon modes and crystalline quality. The primitive unit cell of β-Ga$_2$O$_3$ consists of 10 atoms, resulting in 30 phonon modes, 27 of which are optical [54]. At the Γ-point, these modes belong to the irreducible representation $\Gamma^{opt}$ = 10A$_g$ + 5B$_g$ + 4A$_u$ + 8B$_u$, where the A$_g$ and B$_g$ modes are Raman active, and the A$_u$ and B$_u$ modes are infrared active. For excitation on the (010) plane, only A$_g$ modes are observed in Raman spectra, as B$_g$ modes are forbidden due to the symmetry of the crystal structure. The Raman spectra of the homoepitaxial (010) β-Ga$_2$O$_3$ films as shown in Figure 4(a) exhibits all A$_g$ phonon modes, confirming the (010) orientation of the crystal. The peaks corresponding to A$_g$ modes appear at 109.81, 169.8, 199.18, 319.67, 346.49, 415.84, 474.1, 629.31, 658.64, and 766.50 cm$^{−1}$, which align well with previously reported experimental and theoretical studies on β-Ga$_2$O$_3$ [36,38,54]. These modes represent a range of vibrational behaviors, including low-frequency modes (109.81, 169.8, and 199.18 cm$^{−1}$) related to the vibration and translation of tetrahedra-octahedra chains, mid-frequency modes (319.67, 346.49, 415.84, and 474.1 cm$^{−1}$) linked to Ga$_2$O$_6$ octahedra deformations, and high-frequency modes (629.31, 658.64, and 766.50 cm$^{−1}$) representing the bending and stretching of GaO$_4$ tetrahedra [38, 55-60]. Similarly, the Raman spectra of heteroepitaxial ($\bar{2}$01) β-Ga$_2$O$_3$ films grown on c-plane sapphire substrates are shown in Figure 4(b). Peaks at 114.14, 144.26, 474.1, and 652.45 cm$^{−1}$ correspond to B$_g$ vibrational modes characteristic of the ($\bar{2}$01) orientation, while

additional peaks at 169.27, 200.08, 319.32, 347.1, 416.2, 630.5, and 766.7 cm$^{-1}$ correspond to $A_g$ vibrational modes. The Raman spectra, exhibiting high-intensity and well-defined $A_g$ and $B_g$ modes, confirm the high crystalline quality and phase purity of both homo- and hetero-epitaxial β-$Ga_2O_3$ thin films, with no evidence of other $Ga_2O_3$ phases.

The transport characteristics of β-$Ga_2O_3$ films are also investigated using Hall measurement. By systematically adjusting the $SiCl_4$ flow, the carrier concentrations were effectively tuned. Figure 5 shows the RT mobility vs. carrier concentration for the films grown on both (010) β-$Ga_2O_3$ and off-axis c-sapphire with 6° miscut substrates. A clear correlation between doping concentration and Hall mobility was observed, with higher mobility at lower carrier concentrations. In pure β-$Ga_2O_3$ materials, the room-temperature mobility is typically limited by polar phonon scattering due to strong ionic Ga-O bonding. However, as doping increases, ionized impurity scattering, and neutral impurity scattering become dominant factors limiting the mobility. A similar trend was observed for ($\bar{2}$01) β-$Ga_2O_3$ films grown on sapphire substrates, where doping concentrations between $1.15 \times 10^{17}$ and $8.29 \times 10^{18}$ cm$^{-3}$ were achieved. Table 1 provides a comprehensive summary of the films, detailing their thickness, growth rates, electron mobilities, and carrier concentrations. Both homoepitaxial and heteroepitaxial films exhibited high electron mobilities, with the maximum mobility of 162 cm²/Vs recorded for homoepitaxial films at a doping concentration of $1.51 \times 10^{17}$ cm$^{-3}$ - representing the highest reported mobility for LPCVD-grown homoepitaxial films. Similarly, heteroepitaxial β-$Ga_2O_3$ films grown on c-plane sapphire with a 6˚ offcut exhibited a record-high mobility of 149 cm²/Vs at a doping concentration of $1.15 \times 10^{17}$ cm$^{-3}$. Figure 5 also compares the RT mobility vs. carrier concentration of the films from this work to those grown by other techniques such as MOCVD [22-27,61-67], LPCVD [35-37], HVPE [20, 21], PLD [68], and MBE [41-46]. The results from this study demonstrate the promising mobilities achieved using the LPCVD system, with much faster growth rates (4.7-17 μm/hr) and thicker films (2.3-11.7 μm), highlighting the potential of LPCVD growth system for producing high-quality thick β-Ga₂O₃ films with higher growth rates.

To better understand the charge carrier transport properties of LPCVD grown β-Ga₂O₃ films, temperature-dependent Hall measurements were also conducted on both homoepitaxial and heteroepitaxial thin films. Figures 6(a) and (b) display the temperature dependence of Hall mobility and carrier concentration for (010) β-Ga₂O₃ and ($\bar{2}$01) β-Ga₂O₃ films. The (010) β-Ga₂O₃ homoepitaxial films exhibited RT carrier concentrations of $5.05 \times 10^{17}$ cm$^{-3}$ and $8.29 \times 10^{17}$ cm$^{-3}$ with corresponding RT mobilities of 118 and 106 cm²/V·s, while the ($\bar{2}$01) β-Ga₂O₃ films showed RT carrier concentrations of $1.15 \times 10^{17}$ cm$^{-3}$ and $2.25 \times 10^{17}$ cm$^{-3}$, with corresponding mobilities of 149 and 124 cm²/V·s, respectively. For the two (010) β-Ga₂O₃ films, low temperature peak mobilities of 332 cm²/V·s and 219 cm²/V·s was achieved at 96K and 112K temperatures, respectively with carrier concentrations of $1.55 \times 10^{17}$ cm$^{-3}$ and $3.86 \times 10^{17}$ cm$^{-3}$ as shown in Figure 6(a). Similarly, a record-high low-temperature peak mobilities from the ($\bar{2}$01) β-Ga₂O₃ films exceeding > 843 cm²/V·s and 498 cm²/V·s were observed at ~80 K temperature with carrier concentrations of $1.74 \times 10^{16}$ cm$^{-3}$ and $3.24 \times 10^{16}$ cm$^{-3}$, respectively, indicating the high purity and low compensating defect levels in the ($\bar{2}$01) β-Ga₂O₃ films. All these films exhibited mobility enhancement as the temperature decreases. Notably, electron mobility in β-Ga₂O₃ is primarily limited by ionized impurity scattering at low temperatures and optical phonon scattering at higher temperatures. The donor activation energy ($E_D$) with donor ($N_D$) and compensation ($N_A$) concentrations were extracted by fitting the temperature dependent transport data using the charge neutrality equation ($n + N_A = \frac{N_D}{1 + 2e^{-(E_D - E_F)/k_B T}}$), where $N_A$ is the concentration of acceptors acting as compensators, $N_D$ is the donor concentration, and $E_D$ is the donor activation energy. The extracted activation energies of 13.2 and 7 meV are calculated for the two (010) β-Ga$_2$O$_3$ homoepitaxial films with donor concentrations of $8.2 \times 10^{17}$ cm$^{-3}$ and $1.5 \times 10^{18}$ cm$^{-3}$, respectively with corresponding $N_A$ of $\sim 3 \times 10^{15}$ cm$^{-3}$ and $\sim 5 \times 10^{15}$ cm$^{-3}$. A similar donor activation energy of β-Ga$_2$O$_3$ has been previously reported at comparable doping levels [69]. The donor activation energies reduce as the donor concentration increase, which is consistent with the findings from β-Ga$_2$O$_3$ films investigated with different carrier concentrations [69]. Such decrease in donor activation energies is expected for highly doped semiconductors when an impurity band begins to form [70,71]. Similar trend of a reduction of donor activation energy with the increase in Si doping concentration was also observed in highly Si-doped ($5.2 \times 10^{18}$ cm$^{-3}$ < $N_d$ < $1.5 \times 10^{19}$ cm$^{-3}$) AlGaN films [72]. as well as Si doped β-(Al$_x$Ga$_{1-x}$)$_2$O$_3$ films [73]. In case of ($\bar{2}$01) β-Ga$_2$O$_3$ films, donor concentrations of $1.39 \times 10^{17}$ cm$^{-3}$ and $3.1 \times 10^{17}$ cm$^{-3}$ with corresponding $N_A$ of $\sim 2 \times 10^{15}$ cm$^{-3}$ and $\sim 5 \times 10^{15}$ cm$^{-3}$ were extracted with donor activation energies of 24 and 22 meV, respectively, aligning with previously reported values [36].

To further corelate the SiCl$_4$ flow rates with doping concentration, the C-V measurements were conducted on ~ 6.8 μm thick homoepitaxial β-Ga$_2$O$_3$ films grown on Sn doped (010) β-Ga$_2$O$_3$ substrates, as shown in Figure 7. A schematic cross section of the Schottky diode structure used for this measurement is shown in the inset of Figure 7(a). A monotonous increase in capacitance is observed with the increase in SiCl$_4$ flow rates. The net charge density ($N_D^+ - N_A^-$) values of $8.49 \times 10^{17}$ cm$^{-3}$, $1.11 \times 10^{18}$ cm$^{-3}$ and $2.44 \times 10^{18}$ cm$^{-3}$ are extracted from the C-V charge profile and are plotted in Figure 7(b) for SiCl$_4$ flow rates of 0.07, 08 and 0.10 SCCM, respectively. These results indicate that increasing SiCl$_4$ flow rates leads to higher carrier concentrations, with the $N_D^+ - N_A^-$ values remaining nearly uniform across the depth profile.

To evaluate impurity concentrations, SIMS characterization was conducted on a representative homoepitaxial (010) β-Ga$_2$O$_3$ film with the highest recorded mobility of 162 cm²/Vs, as shown in Figure 8. The analysis revealed a uniform Si distribution throughout the 3.4 μm-thick epilayer, with an average concentration of approximately ~ 4.5 × 10$^{17}$ cm$^{-3}$, exceeding the carrier concentration of 1.51 × 10$^{17}$ cm$^{-3}$ measured by Hall, indicating low activation ratio of Si dopants in the sample. These corroborates well with recent findings from deep level defects investigation in LPCVD grown (010) β-Ga$_2$O$_3$ [39] and can be attributed to the compensation effects caused by native defects unique to the LPCVD growth process. That study using deep-level optical spectroscopy (DLOS) on LPCVD-grown β-Ga$_2$O$_3$ have revealed a dominant defect at E$_C$ - 3.6 eV [39]. This defect, which has not been reported in β-Ga$_2$O$_3$ grown by other methods such as MOCVD and MBE, is a prominent feature in the LPCVD defect spectrum and is believed to act as a strong acceptor-like trap center within the bandgap with concentrations in the mid-10$^{15}$ cm$^{-3}$ range [39]. Further support for this observation comes from DFT calculations, which predict that oxygen vacancies ($V_{O(III)}^+$) or hydrogenated gallium vacancies (V$_{Ga}$ - nH) could create energy levels near E$_C$ - 3.6 eV [74,75]. Although the Si donor ionization energy of < 40 meV in β-Ga$_2$O$_3$ suggests full ionization at room temperature [39,69], the presence of these defects likely trap free carriers, effectively compensating the Si donors and reducing the net carrier concentration. A distinct Si peak was observed at the epilayer/substrate interface in Figure 8, a common feature likely due to surface contamination or residue from substrate processing, as noted in previous studies by LPCVD [35,37], MOCVD [22,23], HVPE [76] and MBE [43,44]. Additionally, carbon and hydrogen concentrations in the films were also investigated by quantitative SIMS, since they can originate from the growth environment, precursors, or instrument components. However, the SIMS results revealed that both carbon and hydrogen levels are below their detection limits. SIMS analysis also confirmed that the Cl concentration is also below the detection limit, indicating the suitability of using SiCl$_4$ as a n-type doping precursor in LPCVD system for achieving high-purity films with controllable low doping.

In summary, LPCVD growth of Si doped β-Ga$_2$O$_3$ films with promising electron transport characteristics have been demonstrated on both (010) Ga$_2$O$_3$ and off-axis c-sapphire substrates. The surface morphology of the films showed a strong dependence on carrier concentrations, with lower doping resulting in smoother surfaces and higher electron mobility. Surface AFM analysis revealed RMS roughness values as low as 2.65 nm for homoepitaxial films, and ranging between 4.29-5.14 nm for heteroepitaxial β-Ga$_2$O$_3$ films. High resolution XRD and Raman measurement confirmed the high crystalline quality and phase purity of the films, with distinct A$_g$ and B$_g$ phonon modes in Raman spectra and no additional phases detected in XRD. Hall measurements showed promising electron mobilities, with a highest RT mobility of 162 cm²/Vs for homoepitaxial films and 149 cm²/Vs for heteroepitaxial films at corresponding RT carrier concentrations of $1.51 \times 10^{17}$ cm$^{-3}$ and $1.15 \times 10^{17}$ cm$^{-3}$ respectively, while temperature-dependent measurements revealed peak mobilities exceeding 843 cm²/Vs at low temperatures (80 K) for the heteroepitaxial film. Donor activation energies ranged from 7 to 24 meV, increasing as doping concentrations decreased. SIMS analysis further confirmed uniform Si distribution and ultra-low impurity concentrations, highlighting high-purity of LPCVD grown β-Ga$_2$O$_3$ films. These results highlight the potential of LPCVD growth system for producing high-quality, thick β-Ga$_2$O$_3$ films with promising transport properties.

**Data Availability**

The data that support the findings of this study are available from the corresponding author upon reasonable request.

**Table 1**

Summary of film thickness, growth rates and transport characteristics of homo- and hetero-epitaxial β-Ga$_2$O$_3$ films.

| Substrate | Estimated Thickness (μm) | Growth Rate (μm/hr) | Hall Mobility (cm$^2$/Vs) | Carrier Concentration (cm$^{-3}$) |
|---|---|---|---|---|
| Fe doped (010) β-Ga$_2$O$_3$ | 3.87 | 7.74 | 118 | 5.05×10$^{17}$ |
| | 3.40 | 6.80 | 162 | 1.51×10$^{17}$ |
| | 3.89 | 7.78 | 92 | 1.41×10$^{18}$ |
| | 3.37 | 6.74 | 101 | 2.25×10$^{18}$ |
| | 4.78 | 9.56 | 106 | 8.29×10$^{17}$ |
| | 3.77 | 7.54 | 81 | 1.19×10$^{19}$ |
| | 2.34 | 4.68 | 113 | 1.25×10$^{18}$ |
| C-plane sapphire with 6˚ offcut | 11.10 | 11.10 | 138 | 2.99×10$^{17}$ |
| | 10.15 | 10.15 | 149 | 1.15×10$^{17}$ |
| | 9.20 | 9.20 | 124 | 2.22×10$^{17}$ |
| | 6.09 | 12.18 | 118 | 3.26×10$^{17}$ |
| | 11.70 | 11.70 | 109 | 3.68×10$^{17}$ |
| | 11.47 | 11.47 | 81 | 1.46×10$^{18}$ |
| | 8.49 | 16.98 | 94 | 5.59×10$^{17}$ |
| | 5.16 | 10.32 | 74 | 1.70×10$^{18}$ |
| | 10.31 | 10.31 | 59 | 8.29×10$^{18}$ |

**Figure Captions**

**Figure 1.** Surface view of FESEM images of β-$Ga_2O_3$ films grown on (a-c) (010) β-$Ga_2O_3$ and (d-f) 6˚ off-axis c-plane sapphire substrates with different carrier concentrations.

**Figure 2.** AFM surface images of β-$Ga_2O_3$ films grown on (a-b) (010) β-$Ga_2O_3$ and (c-d) 6˚ off-axis c-plane sapphire substrates, indicating their corresponding surface RMS roughness.

**Figure 3.** XRD rocking curves for β-$Ga_2O_3$ films: (a) homoepitaxially grown films on (010) β-$Ga_2O_3$ substrate, inset shows a wide ω-2θ scan indicating (020) β-$Ga_2O_3$ peak (b) hetero-epitaxially grown films on c-plane sapphire with inset showing ($\bar{4}$02) and ($\bar{8}$04) peaks of ($\bar{2}$01) oriented β-$Ga_2O_3$ films from ω-2θ scan.

**Figure 4.** Room temperature Raman spectra of (a) homo- and (b) hetero-epitaxial β-$Ga_2O_3$ films.

**Figure 5.** Benchmarking plot of the measured transport properties of β-$Ga_2O_3$ films (RT Hall mobility vs carrier concentration) obtained in this work with values reported in the literature for various growth techniques, including LPCVD [35-37], MOCVD [22-27,61-67], HVPE [20,21], MBE [41-46] and PLD [68].

**Figure 6.** (a) Temperature dependent carrier mobility as a function of temperature for both homo- and hetero-epitaxial films grown with various doping concentrations. (b) Measured carrier concentration as a function of 1000/T and the calculated concentration from the charge neutrality equations. Colored circles represent experimental data points and dashed lines are obtained from the calculation.

**Figure 7:** (a) C-V characteristics of β-$Ga_2O_3$ Schottky barrier diodes fabricated using LPCVD grown β-$Ga_2O_3$ (010) drift layers with three different $SiCl_4$ flow rates. (b) Net carrier concentration profile for all three samples derived from the C-V curves, showing flat charge profile.

**Figure 8:** Quantitative SIMS impurity (Si, C, H and Cl) depth profile of (010) β-$Ga_2O_3$ homoepitaxial film.

**Figure 1**

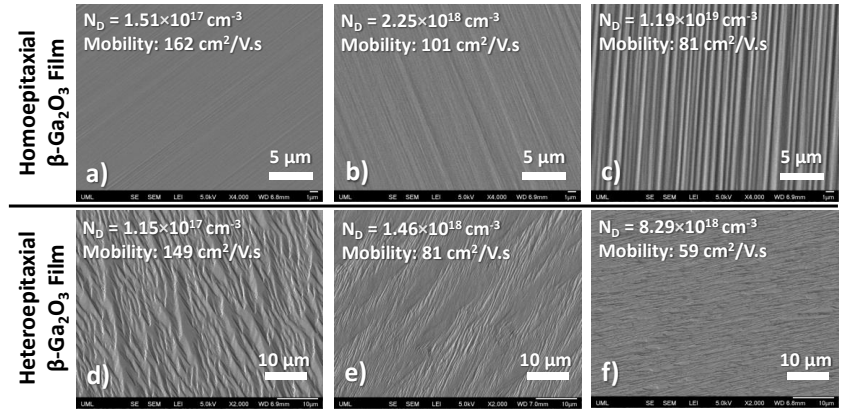

**Figure 2**

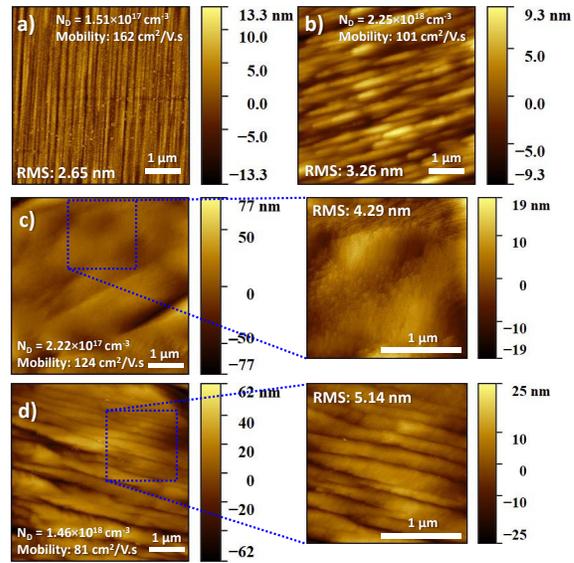

**Figure 3**

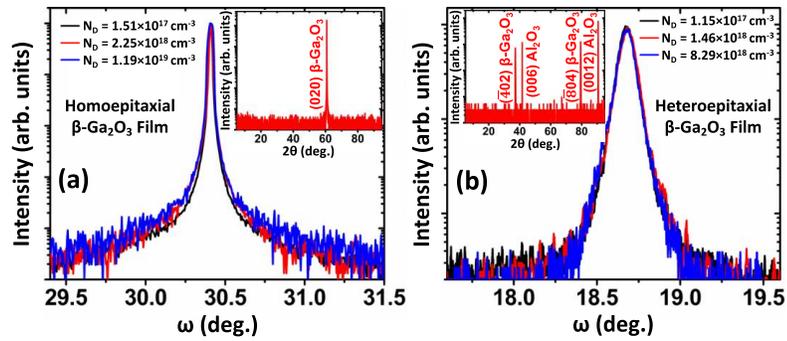

**Figure 4**

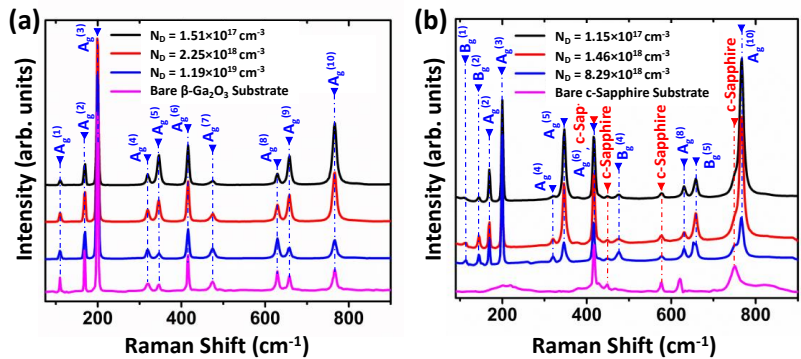

**Figure 5**

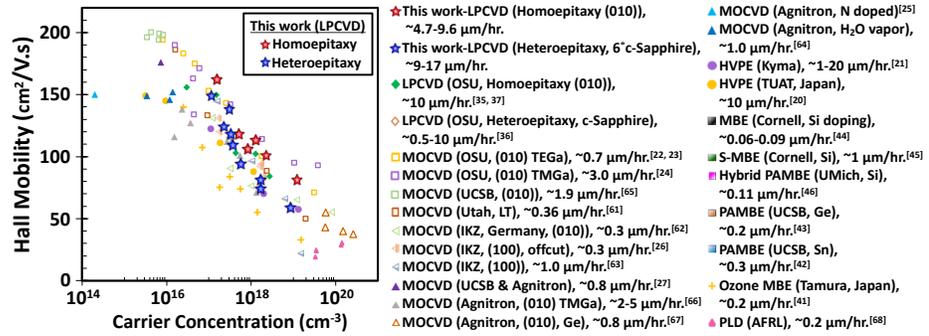

**Figure 6**

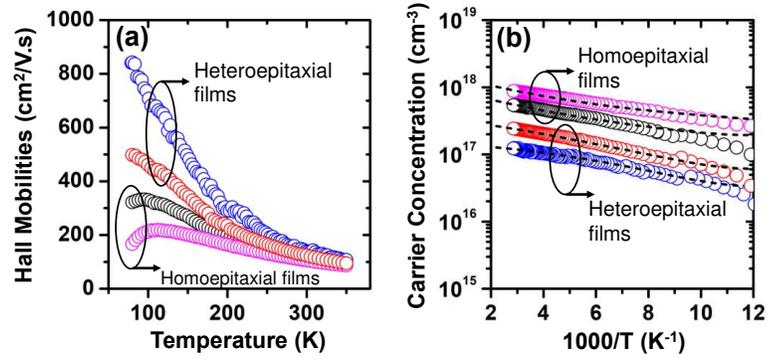

**Figure 7**

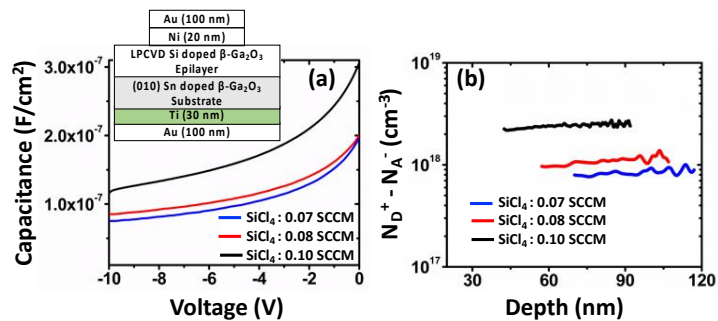

**Figure 8**

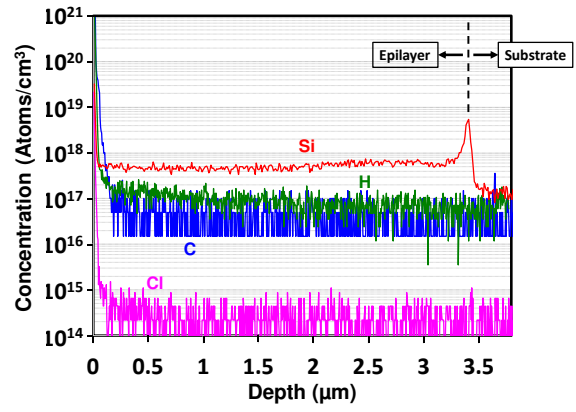